\documentclass{emulateapj}
 
\usepackage{bm}
\slugcomment{\scriptsize{Accepted for publication in ApJ, Preprint typeset using \LaTeX style}} 
\shorttitle{MRI-driven Turbulence in Core-Collapse Supernovae}
\shortauthors{Masada, Takiwaki, Kotake and Sano}
\begin{document}
\title{Local Simulations of the Magneto-rotational Instability in Core-Collapse Supernovae}
\author{Youhei Masada\altaffilmark{1}, Tomoya Takiwaki\altaffilmark{2}, Kei Kotake\altaffilmark{2,3}, and 
Takayoshi Sano\altaffilmark{4}} 
\altaffiltext{1}{Department of Computational Science, Graduate School of System Informatics, Kobe University; 
Nada, Kobe 657-8501:E-mail: ymasada@harbor.kobe-u.ac.jp}
\altaffiltext{2}{Center for Computational Astrophysics, National Astronomical Observatory of Japan,
2-21-1 Osawa, Mitaka, Tokyo, 181-8588, Japan}
\altaffiltext{3}{Division of Theoretical Astronomy, National Astronomical Observatory of Japan, 2-21-1, 
Osawa, Mitaka, Tokyo, 181-8588, Japan: E-mail:kkotake@th.nao.ac.jp}
\altaffiltext{4}{Institute of Laser Engineering, Osaka University, 1-1, Yamadaoka, Suita, Japan: E-mail:sano@ile.ac.jp}

\begin{abstract}
Bearing in mind the application to core-collapse supernovae, we study nonlinear properties of the 
magneto-rotational instability (MRI) by means of three-dimensional simulations in the framework of a local 
shearing box approximation. By changing systematically the shear rates that symbolize the degree of 
differential rotation in nascent proto-neutron stars (PNSs), we derive a scaling relation between the turbulent 
stress sustained by the MRI and the shear-vorticity ratio. Our parametric survey shows a power-law scaling 
between the turbulent stress ($\langle\langle w_{\rm tot} \rangle\rangle$) and the shear-vorticity ratio 
($g_q$) as $\langle\langle w_{\rm tot} \rangle\rangle \propto g_q^{\delta}$ with its index 
$\delta \sim 0.5$. The MRI-amplified magnetic energy has a similar scaling 
relative to the turbulent stress, while the Maxwell stress has slightly smaller power-law index ($\sim 0.36$). 
By modeling the effect of viscous heating rates due to the MRI turbulence, we show that the stronger magnetic 
fields or the larger shear rates initially imposed lead to the higher dissipation rates. For a rapidly rotating PNS 
with the spin period in milliseconds and with strong magnetic fields of $10^{15}$ G, the energy dissipation rate is 
estimated to exceed $10^{51} {\rm erg\ sec^{-1}}$. Our results suggest that the conventional magnetohydrodynamic 
(MHD) mechanism of core-collapse supernovae is likely to be affected by the MRI-driven turbulence, which we 
speculate, on one hand, could harm the MHD-driven explosions due to the dissipation of the shear rotational 
energy at the PNS surface, on the other hand the energy deposition there might be potentially favorable for 
the working of the neutrino-heating mechanism.
\end{abstract}
\keywords{instabilities---rotation--- supernova: magnetic fields --- stars: proto-neutron stars}
\section{Introduction}
Numerical simulations of magnetohydrodynamic (MHD) stellar explosions have started already in the early 
1970's shortly after the discovery of pulsars \citep{leblanc,bisno76,muller79,symb84}. However, it is rather 
only recently that the MHD studies come back to the front-end topics in the supernova research followed by 
a number of extensive MHD simulations (e.g., \citealt{arde00,yama04,kota04a,kota04b,ober06a,ober06b, 
burr07,cerd07,taki09,scheid,taki_kota,martin11}, and \citet{kota06,kotake12,kotake12b} for recent
 reviews). Main reasons for this activity are observations 
indicating very asymmetric explosions \citep{wang01,wang02}, and the interpretation of magnetars \citep
{dunc92,latt07} and gamma-ray bursts (e.g., \citealt{woos06,yoon}) as a possible outcome of the 
magnetorotational core-collapse of massive stars.
 
The MHD mechanism of stellar explosions relies on the extraction of rotational free energy of collapsing 
progenitor core via magnetic fields. Hence a high angular momentum of the core is preconditioned for 
facilitating the mechanism \citep{meier}. Given (a rapid) rotation of the  pre-collapse core, there are at least 
two ways to amplify the initial magnetic fields to a dynamically important strength, namely by the field 
wrapping by means of differential rotation that naturally develops in the collapsing core, and by the so-called 
magneto-rotational instability (MRI, see \citealt{balb98}). 
 
\citet{akiy03} were the first to point out that the interfaces surrounding the nascent protoneutron stars (PNSs) quite 
generally satisfy the instability criteria for the MRI. Therefore any seed magnetic fields can be amplified exponentially 
in the differentially rotating layers, much faster than the linear amplification due to the field wrapping. After the MRI 
enters to the saturated state, the field strength might reach $\sim 10^{15-16}$ G, which is high enough to affect 
the supernova dynamics. Not only to amplify the magnetic fields, the MRI plays a crucial role also in operating 
the MHD turbulence (see \citealt{hawley95,balb98,masada06}). The turbulent viscosity sustained by the MRI 
can convert a fraction of the shear rotational energy to the thermal energy of the system. \citet{thomp05} 
suggested that the additional energy input by turbulent viscous heating can help the neutrino-driven supernova 
explosion. Followed by the exponential field amplification and the additional heating, a natural outcome of the 
magnetorotational core-collapse may be the formation of energetic bipolar explosions, which might be observed 
as the so-called hypernovae (see \citealt{tanaka} and references therein).

Here it is noted that bipolar explosions obtained in the previous MHD supernova simulations mentioned above are 
not driven by the MRI,  but predominantly by the field wrapping, assuming very strong pre-collapse magnetic fields 
($ \gtrsim 10^{12}$G) in general. The growth rate of MRI-unstable modes depends on the product of the initial field 
strength and the wavenumber of the mode. In the case of the canonical initial fields ($\sim 10^{9}$G) as predicted 
by recent stellar evolution models \citep{hege05}, the fastest growing modes are estimated to be at most a few 
meters in the collapsing iron core \citep{MRI}. Unfortunately however, it is still computationally too expensive to 
resolve those scales in the global MHD simulations, typically more than two orders-of-magnitudes smaller than 
the typical finest grid size. To reveal the nature of the MRI, local simulations focusing on a small part of the 
MRI-unstable regions are expected to be quite useful as traditionally studied in the context of accretion disks 
(see \citealt{balb98}).

\citet{MRI} were the first to report their numerical simulations to study the linear growth and nonlinear properties 
of the MRI in the supernova environment. To ease a drawback of the local shearing box simulation, they employed 
the shearing disk boundary conditions by which the global radial density in the vicinity of the equatorial region
in the supernova core can be taken into account. By performing such a {\it semi-global} simulation systematically 
in two (2D) and three dimensions (3D), they derived scaling laws for the termination of the linear growth of the MRI. 
As estimated in \citet{akiy03}, the MRI was shown to amplify the seed fields exceeding $10^{15}$G. These important 
findings may open several questions that motivate us to join in this effort, such as how the non-linear properties as 
well as the scaling laws could be changed by other parameters yet unexplored in the supernova context (that we are 
going to explain in the next paragraph), and whether the viscous heating maintained by the MRI turbulence could or 
could notaffect the supernova mechanism. 

A fundamental and long-lasting issue regarding the MRI itself is to understand the features of the MRI in the non-linear 
phase and to specify which physical quantities determine the saturation levels of the MRI. So far, extensive efforts have 
been paid to measure parameter dependence of the MRI-sustained turbulence by means of numerical simulations.
It was reported that the amplitude of the turbulent stress maintained by the MRI depends on the net value of the initial 
magnetic field and the gas pressure of the system, that is
\begin{equation}
w_{\rm tot} \propto B^{\xi} P^{\zeta}\;,
\end{equation}
where $\xi$ and $\zeta$ are the power law indices (see also Blackman et al. 2007 for the scaling relation between 
the turbulent stress and the plasma beta).  Hawley et al. (1995) found that the saturation amplitude depends on the field 
strength when the system is penetrated initially by a uniform vertical field ($w_{\rm tot} \propto B$), but is
independent of the initial field strength if there is no net magnetic flux in the system (see also Sano et al. 2004). However, 
in the density stratified system,  it is recently suggested that the turbulent stress increases almost linearly with magnetic 
energy of the net vertical field ($w_{\rm tot} \propto B^2$; Suzuki et al. 2010; Okuzumi \& Hirose 2011). A weak 
dependence of the gas pressure on the MRI turbulence was found by Sano et al. (2004):  $\zeta \simeq 1/4$ for models 
with initial net-zero magnetic flux and $\zeta \simeq 1/6$ in the system penetrated by uniform magnetic field. The physical 
mechanisms responsible for these parameterdependences remain an issue under considerable debate.

In the stellar interior condition, such as in the supernova core, it should be important to understand how the degree of 
differential rotation of the PNS could affect the nonlinear properties of the MRI. This is because the shear rate at the 
PNS surface ($q \equiv -d \ln \Omega/d \ln r$ with $\Omega$ and $r$ representing the angular velocity and radius, 
respectively) is time- and location- dependent unlike accretion disks in which a force balance between the centrifugal 
force and gravity is generally maintained.

In this work, we study the MRI-driven turbulence and their nonlinear properties by performing three-dimensional 
simulations in the framework of a local shearing box approximation. By changing systematically the shear rates that 
symbolize the degree of differential rotation in nascent PNSs, we specifically study how the nonlinear properties 
of the MRI-driven turbulence respond to it. Based on our numerical results, we estimate the energy deposition rate 
due to the viscous heating driven by the MRI and discuss its potential impacts on the supernova mechanism. 

This paper opens with descriptions of the numerical methods and the initial settings in \S~2. The numerical results 
are presented in \S~3. Based on the numerical results, we then move on to address possible impacts of the MRI on 
the supernova mechanism in \S~4 followed by discussions in \S~5. We summarize our new findings in \S~6. 
\section{Numerical Methods and Initial Settings}
To study nonlinear properties of MRI, MHD equations are solved with a MPI-parallelized finite-difference code that 
was originally developed by \citet{sano98}. The hydrodynamic part is based on the second-order Godunov scheme 
\citep{leer}, which consists of Lagrangian and remap steps. The exact Riemann solver is modified to account for the 
effect of tangential magnetic fields. The field evolution is calculated with Consistent MoC-CT method and thus can 
avoid the numerical explosive instability appeared in the strong shear layer when the MoC method is adopted \citep{clarke}. 
The energy equation is solved in the conservative form. The advantages of our scheme are its robustness for strong 
shocks and the satisfaction of the divergence-free constraint of magnetic fields 
\citep{evans,stone}.

We perform a series of 3D compressible MHD simulation by adopting the local shearing box model described 
in detail by \citet{hawley95} . In the shearing box model, MHD equations are written in a local Cartesian frame 
of reference $(x,\ y,\ z)$ co-rotating with the local portion of the stellar interior rotating at the angular 
velocity $\Omega$ corresponding to a fiducial radius in the cylindrical coordinate $R$. Then the coordinates 
are presented as $x = r - R $, $y = R\phi - \Omega t$, and $z$. The fundamental equations are written in terms 
of these coordinates within a small region surrounding the fiducial radius, in $\Delta r \ll R $, 
\begin{eqnarray}
\frac{\partial \rho}{\partial t} + \nabla\cdot (\rho \bm{v})   &=&  0 \;, \label{eq1} \\ 
\frac{{\partial }\bm{v}}{ \partial  t} + \bm{v}\cdot \nabla \bm{v}  & = & - \frac{1}{\rho} \nabla 
\left( P + \frac{| \bm{B}|^2}{8\pi} \right)+ \frac{(\bm{B}\cdot \nabla)\bm{B}}{4\pi\rho }  \nonumber \\ 
&& - 2\Omega\times \bm{v} - 2q\Omega^2 x {\tilde{\bm{x}}}   \;, \label{eq2} \\
\frac{\partial \epsilon }{\partial t} + (\bm{v}\cdot\nabla) \epsilon & = & - \frac{P}{\rho} \nabla\cdot \bm{v} 
+ \Phi \;, \label{eq3}\\
\frac{\partial \bm{B} }{\partial t} &=& \nabla \times (\bm{v} \times \bm{B} ) \;, \label{eq10}
\end{eqnarray}
where $\epsilon $ is the specific internal energy, and the other parameters have their usual meanings. The 
term $-2q\Omega^2 x$ in the momentum equation is the tidal expansion of the effective potential with a shear 
rate $q \equiv - d\ln\Omega/d\ln r$. Assuming the ideal gas, the pressure is given by 
$P = (\gamma -1)\rho \epsilon $. The constant ratio of specific heats $\gamma$ is considered for simplicity. 
We choose adiabatic gas with $\gamma = 5/3$ in this paper. 

Since we focus on the local properties of the instability, we employ, as is described schematically in Figure~1,  
numerical grid representing a small portion of the convectively stable upper PNS below the neutrinosphere 
where the strong shear is naturally developed when the core-collapse (Akiyama et al. 2003; Thompson et al. 2005). 
Adopting the differentially rotating matter as an unperturbed state, the azimuthal velocity is given by 
$v_y = q \Omega x $ in the frame co-rotating with the velocity $R \Omega$. The radial force balance at the initial 
state is thus achieved between the Coriolis force and the tidal force which is the residual of the gravitational force 
mainly balancing with the background pressure gradient force. We discuss an applicability of our numerical results 
to the high-temperature and highly stratified plasma realized in the supernova environment and the effects of 
neutrino viscosity in \S5. 

We choose normalization with $\rho_0 = 1$, $\Omega = 10^{-3}$, and the computational domain has a 
radial size $L_x = 4$, an azimuthal size $L_y=4$, and a vertical size $L_z = 1$. All of the runs use a 
uniform grid of $256\times 256 \times 64$ zones. The initial field geometry is a uniform vertical magnetic 
field {\boldmath $B$} $= B_0${\boldmath $e_z$} (net non-zero flux). We assume, for all the models, that 
the initial gas pressure is $P_0 = 5\times 10^{-7}$ and the initial ratio of the gas and magnetic pressures, that 
is the plasma beta is $\beta_0 = 3200$. The pressure scale height is then $H_p = 1$, and is the same as the 
vertical box size. With this normalization, the initial field strength is $B_0 = 6.26\times 10^{-5}$, yielding 
$v_{A0} = 1.77\times 10^{-5}$, where  $v_{A0} = B_0/(4\pi\rho_0)^{1/2}$ is the Alfv\'en speed. Note that 
these non-dimensional parameters can be translated to the dimensional form suitable for supernova environments 
as $\rho = 10^{12}\ {\rm g\ cm^{-3}}$, $L = 10^{2}\ {\rm cm}$, $\Omega = 10^2\ {\rm rad\ sec^{-1}}$, 
$P = 5 \times 10^{25}\ {\rm dyn\ cm^{-2}}$,  and $B = 2.2 \times 10^{12}\ {\rm G}$, which are summarized 
in Figure~1. 
 
The initial characteristic wavelength of the MRI is given by 
$\lambda_{\rm MRI} = 4\pi\eta v_{Az0}/\Omega $ for ideal MHD case \citep{balb98}, where 
$\eta \equiv 1/\sqrt{4q-q^2}$. Its ratio to the vertical box size is thus varied as 
$\lambda_{\rm MRI}/L_z = 0.22\eta $ in our models. It is empirically known that, to gain a saturation 
level correctly, the MRI wavelength must be resolved at the saturated state by at least six grid zones 
(see Sano et al. 2004). Here we choose the initial setting which satisfies this empirical rule for all the
 models. 
 
\section{Numerical Results}
\subsection{A Fiducial Run}
As a fiducial run, we examine temporal evolutions of the MRI in a model with a shear rate $q = 0.5$. 
Initial perturbations are introduced by giving a random velocity perturbations which are taken to 
have a zero mean value with a maximum amplitude of $|\delta v|/\sqrt{\gamma P_0/\rho_0} = 5 \times 10^{-3}$. 

To characterize the properties of the MRI-driven turbulence, we pay particular attention to the magnetic 
energy, the Maxwell and Reynolds stresses, and total turbulent stress defined respectively by,
\begin{eqnarray}
E_{\rm mag} &\equiv& |{\boldmath B}|^2/8\pi, \\
w_{M} & \equiv & - B_rB_\phi /4\pi, \\
w_{R} & \equiv & \rho v_r \delta v_\phi, \\
w_{\rm tot} & \equiv & w_{M} + w_{R}  \;,
\end{eqnarray} 
where $\delta v_\phi$ is the perturbed azimuthal velocity. Note that a volume average of these quantities is 
represented by single brackets (like $\langle E_{\rm mag} \rangle$) and a time- and volume-average is by 
double brackets (like $\langle \langle E_{\rm mag} \rangle \rangle$). 

The thick, dash-dotted and dashed curves in Figure~2 show the temporal evolution of volume-averaged 
magnetic energy, Maxwell and Reynolds stresses normalized by the initial gas pressure $P_0$. Note that the 
horizontal axis is normalized by the rotation period $t_{\rm rot} \equiv 2\pi/\Omega = 6.28\times 10^3$ 
$(62.8\ {\rm [ms]}$ in dimensional form). In order for the demonstration of the typical evolution of the MRI, 
four time snapshots of the 3D structure of the radial magnetic field are visualized by means of volume rendering 
method in Figure~3. The color bar indicates the amplitude of the radial magnetic field, yellow for the positive values, 
and blue for the negative value. Panels (a)--(d) correspond to the snapshots at the times $t = 6t_{\rm rot}$, 
$8t_{\rm rot}$, $9t_{\rm rot}$, and $11t_{\rm rot}$ respectively. 

As was described in the previous MRI studies for accretion disks with Keplerian rotation of $q = 1.5$, 
there are three typical evolutionary stages observed in our fiducial run, those are (i) linear exponential growth 
stage, (ii) transition stage, and (iii) nonlinear turbulent stage \citep{hawley92,hawley95}. Each stage is denoted 
by white, dark gray and light gray shaded regions in Figure~2. 
 
In the stage (i), the channel structure of the magnetic field exponentially evolves and inversely cascades to the 
larger spatial scale with small structures merging as the magnetic field is amplified. This is because the channel 
mode of the MRI is an exact solution even for the nonlinear MHD equation \citep{goodman}. The temporal 
evolution of channel structures for the radial magnetic field is shown in the panel (a) of Figure~3. 

Then in the stage~(ii), the channel structure of the magnetic field is disrupted via the parasitic instability and/or 
magnetic reconnection at the transition stage \citep{goodman,MRI}. The channel disruption induces a drastic 
phase-shift from the coherent structure to the turbulent tangled structure of the field as is illustrated by the panels 
(b) and (c) of Figure~3. The magnetic energy stored in the amplified magnetic field is then converted to the 
thermal energy of the system. 

Finally in the stage~(iii), the nonlinear turbulent state emerges after the channel disruption is maintained 
long enough by the MRI-driven turbulence as is demonstrated in the panel (d) of Figure~3. Figure~2 shows 
that, at this stage, the turbulent Maxwell stress dominates the Reynolds stress. It then converts the free energy 
stored in the differential rotation into the thermal energy of the system. The time- and volume-averaged total 
turbulent stress is saturated, in the fiducial model, at the level 
\begin{equation}
\langle\langle w_{\rm tot} \rangle\rangle \simeq 0.03P_0 \;, \label{eq_fid}
\end{equation}
where the time average is taken over the saturated state from $50t_{\rm rot}$ to $150t_{\rm rot}$. It can be 
stated that the variation of the shear rate does not change the nonlinear properties of the MRI qualitatively. 

In the shearing box model adopted in our study, such free energy is continuously injected from the radial 
boundaries. This is the reason why the turbulent stage powered by the MRI is maintained in our simulation. 
We discuss in \S~5 the applicability of our numerical model in comparison with the shearing disk model 
employed in \citet{MRI}.
\subsection{Shear Rate Dependence}
In the stellar interior,  such as supernova cores, the rotation profile should be time- and location-dependent 
unlike the typical accretion disk which maintains a quasi-Keplerian rotation with the shear rate $q \simeq1.5$. 
This is because the force balance is mainly achieved between the gravity and the pressure gradient 
force in the stellar interior. It should be important to study how the nonlinear properties of the MRI would 
respond to the change of the shear rate for estimating its impact on the supernova dynamics. The shear 
rates used in our parametric survey are listed in Table 1 with a few diagnostic quantities, the time- and 
volume-averaged Maxwell stress, Reynolds stress, magnetic energy and the ratio of Maxwell and Reynolds 
stresses,  calculated from the numerical data. 

Figure~4 shows the temporal evolution of the turbulent stress for the models with different shear rates.  
The grey, red, blue, orange, green and purple curves indicate the models with $q=0.2$, $0.5$, $0.8$, 
$1.2$, $1.5$ and $1.8$ respectively. Note that the other physical parameters are the same as those used in 
the fiducial run. It is found that both the linear growth rate and saturation amplitude of the turbulent stress 
become larger for the model with the larger shear rate. This trend can be recognized as well from the spatial 
structures of magnetic fields in the saturated state. 

The volume rendered vertical magnetic field is visualized for the models with (a) $q=0.5$ and (b) $q=1.5$ in 
Figure~5. The time snapshot at $t = 100t_{\rm rot}$ is chosen for both models. The red color corresponds to 
the positive value of the vertical field component and the blue is its negative value. The larger structure of the 
magnetic field is found to be generated for the model with the larger shear rate. The other components of the 
magnetic field have similar nonlinear properties. 

Figure~6(a) exhibits the time- and volume-averaged Maxwell stress $\langle\langle w_{M} \rangle\rangle$ 
(red circles), Reynolds stress $\langle\langle w_{R} \rangle\rangle$ (blue squares), total stress $\langle
\langle w_{\rm tot} \rangle\rangle$ (green diamonds), and magnetic energy $\langle\langle E_{\rm mag} 
\rangle\rangle$ (yellow triangles) as a function of parameter $g_q$, where $g_q \equiv q/(2-q) $ is the ratio 
of vorticity $(2-q)\Omega$ to the shear $q\Omega$ (hereafter called ``shear-vorticity ratio'', see Abramowicz et al. 1996). 
Note that we take, at the saturated state, the temporal average of volume-averaged stresses and magnetic energy 
from $50t_{\rm rot}$ to $150t_{\rm rot}$ for models~1,2,3, and from $100t_{\rm rot}$ to $150t_{\rm rot}$ for 
models~4,5,6,7. The black line, which is proportional to $g_q^{1/2}$, is plotted as a reference. 

The Maxwell stress dominates the Reynolds stress for all the models we surveyed although the fraction of the 
Reynolds stress in the total stress increases with the shear-vorticity ratio $g_q$. It should be noted that the pure 
hydrodynamic shear instability develops in the system with the stronger shear of $q \ge 2$ (the so-called Rayliegh 
criterion, e.g., \citealt{chandra}). According to \citet{hawley99}, the hydrodynamic turbulence is shown to change 
the properties of MRI turbulence when the Rayliegh criterion is satisfied (see also \citealt{workman}). The Reynolds 
stress then exceeds the Maxwell stress. However, since $q \lesssim 2$ is generally satisfied in the vicinity of the 
PNSs (e.g., \citealt{kota04a,ober06a,MRI}), the Maxwell stress is expected to dominate over the Reynolds stress 
and plays a major role in the turbulent heating process there. 

A power-law relation between the shear-vorticity ratio and turbulent stress appears in this figure. The power law 
index $\delta$ of a scaling relation defined by 
\begin{equation}
\langle\langle w_{\rm tot} \rangle\rangle \propto  g_q^{\delta} \;, \label{eq_svr}
\end{equation}
is about $1/2$ and the best fit is $\delta = 0.44$ among all the models in Figure~6(a). It would be interesting that 
the magnetic energy has a similar scaling relation to the turbulent stress, that is roughly 
$\langle\langle E_{\rm mag} \rangle\rangle \propto g_q^{1/2}$ (the best fit is $\delta = 0.46$) although the 
Maxwell stress has a power law index of $0.36$ (the best fit value) which is a bit smaller than that for the total stress. 
As was well known that the maximum growth rate of the MRI is linearly proportional to the shear rate $q$ (that is 
$\gamma_{\rm max} = q\Omega/2$), the scaling law of the turbulent stress we found here should be explained 
through the nonlinear properties of the MRI-driven turbulence. 

\subsection{Our Work versus Previous Studies}
Thus far, the dependence of the non-linear properties of the MRI on the shear rate has been investigated 
only in models whose net magnetic flux is set to zero  \citep{brandenburg, abra, hawley99, ziegler, 
pessah}. These previous studies found the stronger dependence of the turbulent stress on the shear-vorticity 
ratio $g_q$ (or rather shear rate $q$ itself) which is in contrast to the weak dependence ($\delta = 1/2$ see 
equation (\ref{eq_svr})) obtained here in the case of the non-zero magnetic flux. 

Abramowicz et al. (1996; hereafter ABL96) numerically evaluated the MRI-sustained turbulent stress for different 
values of the shear-vorticity ratio in the stratified system with net zero magnetic flux. They found that the turbulent 
stress is roughly represented by a linear dependence on the shear-vorticity ratio, that is $\delta = 1.0$ of equation
(\ref{eq_svr}). On the other hand, the numerical results obtained by Ziegler \& R\"udiger (2001; hereafter ZR01) 
with a similar numerical setting as that adopted in ABL96 suggested that the existence of a relation 
$\langle\langle w_{\rm tot} \rangle\rangle \propto q $ [i.e., $\propto 2g_q/(g_q+1)$] in contrast to the finding 
of ABL96. 

The dependence of the turbulent stress on the shear rate was studied comprehensively by Hawley et al. (1999; 
hereafter HBW99) using unstratified shearing box model with net zero magnetic flux . They found that the vorticity 
($(2-q)\Omega$) limits hydrodynamic turbulence and strongly reduces the Reynolds stress at around $q=0$. 
In contrast, the shear ($q\Omega$) promotes turbulence, and thus the Maxwell and Reynolds stresses both are 
enhanced with increasing the shear rate. Since the $q$-dependence of the Reynolds stress is much stronger than 
that of the Maxwell stress, the stress ratio diminishes with increasing $q$. When extracting the data from Figure~10 
of HBW99, the time- and volume-averaged turbulent stress seems to behave as $\langle\langle w_{\rm tot} 
\rangle\rangle \propto g_q^{1/2}$ (the best fit is $\delta = 0.55$). 

Liljestr\"om et al. (2009; hereafter LKKBL09) studied, using local shearing box model with net zero magnetic flux, 
the nonlinear properties of MRI-driven turbulence with varying shear rate. By fitting numerical data listed in Table~1 
of LKKBL09, the turbulent stress has strong $q$-dependence like as that of ABL96 and behaves approximately 
as $\langle\langle w_{\rm tot}\rangle\rangle \propto g_q $ (the best fit is $\delta = 0.9$).

A theoretical model for the saturation of the Maxwell and Reynolds stresses in MRI turbulence was developed by 
Pessah et al. (2006; hereafter PCP06) in light of the similarities exhibited by the linear regime of the MRI and the 
turbulent state. On the basis of linear theory, they formulate a predictor function about the ratio of the Maxwell and 
Reynolds stresses at the saturated state, that is $\langle\langle w_M\rangle\rangle / \langle\langle 
w_R\rangle\rangle = (2+g_q)/g_q\ [= (4-q)/q\ {\rm in\ the\ other\ form}]$. LKKBL09 found that the relation for 
the stress ratio derived by PCP06 predicts similar behavior, but provides a few times smaller ratio than that 
computed from their numerical results. The local closure model developed by Ogilvie (2003) rather could reproduce 
numerical results of LKKBL09 quite well. 

For the comparison between our work and the previous studies, the ratio of the time- and volume-averaged Maxwell and 
Reynolds stresses is plotted as a function of the shear-vorticity ratio $g_q$ in Figure~6(b). The blue squares correspond to 
our result, the green crosses show the numerical data of HBW99, the orange triangles are the result of LKKBL09, and the red 
circles present the result of ZR01. A power law relation of $\langle\langle w_M\rangle\rangle / \langle\langle w_R\rangle\rangle 
\propto g_q^{-3/4}$ is shown by the solid line for the reference. The theoretical model of the stress ratio developed by PCP06 
is traced by the dashed line. Remember that the previous studies, as was summarized above, provided different
 $g_q$-dependences of the turbulent stress. The non-zero magnetic flux is imposed only in our work though the shearing 
 box approximation is adopted in all the models. The stress ratio of HBW99 is calculated by extracting the data from 
 Figure~10 of HBW99. 

The stress ratio decreases with the increase of the shear-vorticity ratio $g_q$ for all the models. This suggests that the Reynolds 
stress has generally a stronger $g_q$-dependence than that of the Maxwell stress.  As LKKBL09 discussed, the model stress 
ratio of PCP06 predicts similar behavior, but gives a few times smaller value than that obtained by all the numerical models 
especially in the range $0.2 \lesssim g_q \lesssim 10$.  It is a bit surprising that the stress ratio is similar among three works 
except HBW99 although they predict different power-law relations between the turbulent stress and the shear-vorticity ratio. 
The stress ratio is roughly represented by a power-law relation $\langle\langle w_M\rangle\rangle 
/ \langle\langle w_R\rangle\rangle \propto g_q^{-3/4}$ in the range of $0.5 \lesssim q \lesssim 1.8$. These similar nonlinear characteristics lied behind the MRI-driven turbulent flows suggest that the stress ratio might be one of key parameters for 
validating the nonlinear saturation mechanism of the MRI. 
\section{Impacts on the Supernova Explosion}
As taken in \citet{thomp05}, the $\alpha$-description \citep{shakura} conventionally used in the context of 
accretion disks is the simplest way to model the turbulent heating. However, it is quite uncertain whether the 
prescription is really applicable to the supernova problem. Based on our numerical results, we hope to 
reexamine whether the turbulent heating due to the MRI could have impacts on the explosion mechanism.

Combining our result with scaling relations of the turbulent stress described by equation (1), the 
heating rate maintained by the MRI turbulence can be expressed as,
\begin{eqnarray}
\epsilon_{\rm MRI}  && =   \langle\langle w_{\rm tot} \rangle\rangle q \Omega \nonumber \\
			          && =  \epsilon_0 \left(\frac{B}{B_0}\right) ^{\xi}\left(\frac{P}{P_0}\right)^{\zeta} \left(\frac{g_q}{g_{q0}}\right)^{\delta} \left(\frac{q}{q_0}\right) \left(\frac{\Omega}{\Omega_0}\right) \;, 
\end{eqnarray}
where $\epsilon_0$ represents the turbulent heating rate at the reference state with $B_0$, $P_0$, $g_{q0}$, 
$q_0$, and $\Omega_0$. The power law indices are positive values depending on the presence of the stratification 
or net magnetic flux, and are expected to be $\xi = 1\sim 2$ (Hawley et al. 1995; Sano et al. 2004; Suzuki et al. 2010; 
Okuzumi \& Hirose 2011), $\zeta = 1/4\sim 1/6$ (Sano et al. 2004), and $\delta = 1/2 \sim 1$ (see \S3.3). 

When choosing the physical parameters adopted in the fiducial model as the reference state, that is 
$B_0 = 2.2 \times 10^{12}\ {\rm G}$, $P_{0} = 5 \times 10^{25}\ {\rm erg}~{\rm cm}^{-3}$, 
$\Omega_{0} = 10^2\ {\rm rad}~{\rm s}^{-1}$, $g_{q0} = 0.33$ and $q = 0.5$, we can estimate, from equation~(\ref{eq_fid}),
the reference heating rate as $\epsilon_0 =  0.03 P_0q_0\Omega \simeq 10^{26} \ {\rm erg\ cm^{-3}\ sec^{-1}}$ 
(see \S3.1 for the fiducial run). 

We make a crude estimate of the volume in which the MRI-driven turbulence becomes active as 
$V_{\rm MRI}  = 4\pi R^2 h\simeq  10^{21}\ {\rm cm^3}R_7^2h_6 $, where $R_7 = R/10^7$ cm is typical radius 
of the PNS normalized by $10^7$ cm, and $h_6 = h/10^{6}$ cm is the typical radial thickness of the MRI-active layer 
normalized by $10^6$ cm. Then the energy releasing rate $L_{\rm MRI} \equiv \epsilon_{\rm MRI} V_{\rm MRI}$ 
becomes, 
\begin{eqnarray}
L_{\rm MRI} & \simeq & 10^{47} R_7^2h_6\ {\rm erg\ sec^{-1}} \times  \label{eq_lmnsty}\\ 
&& \left( \frac{B}{B_0}\right)^\xi \left( \frac{P}{P_0} \right)^{\zeta}
		       \left(\frac{g_q}{g_{q0}}\right)^{\delta} \left(\frac{q}{q_0}\right) \left(\frac{\Omega}{\Omega_0}\right) \;, 
		       \nonumber
\end{eqnarray}
Since we are interested in the MRI in the post-bounce stage of core-collapse supernovae, the gas pressure in the neutrino 
opaque upper PNSs is a known parameter and should be $P = 10^{31}\ {\rm erg\ cm^{-3}}$. When applying the typical
gas pressure at the post-bounce phase to equation~(\ref{eq_lmnsty}) with remaining the other parameters, the energy releasing rate is 
slightly enhanced and becomes $L_{\rm MRI} \simeq 10^{48} \ {\rm erg\ sec^{-1}}$ almost independent of $\zeta$. 
The magnetic field strength and the rotation profile are fully unknown parameters in the supernova environment.  These thus 
determine whether the MRI-sustained turbulent heating can assist the supernova explosion or not. 

The energy releasing rate due to the MRI-driven turbulence $L_{\rm MRI}$ is mapped in Figure~7 as functions of the magnetic 
field strength $B$ and the shear rate $q$. Panels~(a) and (b) show the models with the weakest parameter dependence of 
$\xi =1$ and $\delta = 1/2$, while panels~(c) and (d) are for the models with the strongest parameter dependence of $\xi =2$ and $\delta = 1$. The slow rotation of $\Omega = \Omega_0$ is assumed in left panels and the fast rotation of $\Omega = 10\Omega_0$ 
is in right panels. The green, blue and red dashed curves correspond to the energy releasing rates $10^{50}$, $10^{51}$, and 
$10^{52}\ {\rm erg\ sec^{-1}}$ respectively. The darker color gives the higher energy releasing rate. Note that the gas pressure is 
fixed to be $10^{31}\ {\rm erg\ cm^{-3}}$ and then the weak gas pressure dependence of the MRI-driven turbulence is neglected 
here for the simplicity. 

The dark shaded parameter space provides the larger energy releasing rate than about $10^{51}\ {\rm erg\ sec^{-1}}$ which 
should assist the supernova explosion.\footnote{The typical timescale of neutrino-driven explosions observed in the recent 
supernova simulations is $\gtrsim 300$ ms after bounce (e.g., \citealt{marek,Suwa10,Suwa12,bernhard12_3}). 
So the energy deposition could be as high as $10^{50}~{\rm erg}$ when the MRI-driven turbulence maintains the heating 
of $10^{51}\ {\rm erg\ sec^{-1}}$. Since the explosion energy of core-collapse supernovae is estimated as $10^{51}\ 
{\rm erg}$, the turbulent heating due to the MRI is expected to play an important role in assisting explosion when 
$L_{\rm MRI} \gtrsim 10^{51}\ {\rm erg\ sec^{-1}}$.}. The MRI-turbulent heating then have a minor effect on the supernova 
explosion in the unshaded parameter space which provides the luminosity $L_{\rm MRI} \ll 10^{51}\ {\rm erg\ sec^{-1}}$. 

It is fairly obvious that the stronger magnetic field or the larger shear rate yields the higher energy release due to the MRI, 
and thus makes a greater contribution on the supernova explosion. When considering the moderate shear of $q\simeq 1$
and angular velocity of $\Omega = \mathcal{O}(10^2)\ {\rm rad\ sec^{-1}}$ which are plausible for the supernova environment 
(Ott et al. 2006), a strong magnetic field of $B \gtrsim 10^{15}\ {\rm G}$ is required for the MRI-assisted supernova explosion 
in the case with the weakest parameter dependence of $\xi =1$ and $\delta = 1/2$ [Panels (a) and (b)]. The magnetic field 
weaker than $10^{13}\ {\rm G}$ can drive the MRI-driven turbulence strong enough for assisting the explosion when the 
MRI turbulence depends strongly on the physical parameters $B$ and $q$ [see Panels (c) and (d)]. 

The nonlinear MRI studies employing shearing box model might suggests that non-canonical post-bounce states, such as 
having strong magnetic field and/or large spin rate, must be developed for prompting MRI-assisted supernova explosion. 
It may be tempting to implement those phenomenological heating rates as sub-grid model into the global MHD supernova 
simulations to see the outcomes. However, before that, it is more important to conduct a more extensive nonlinear study of 
the MRI over a wider parameter range and/or using more realistic numerical setting in order to precisely fix the scaling relations. 

\section{Discussion}
\subsection{Effects of Neutrino Viscosity on MRI}
In the neutrino opaque upper PNSs where we have a great interest on in this paper, the neutrino viscosity was 
shown to suppress the growth of the MRI when the viscous dissipation timescale of a typical MRI mode becomes 
shorter than the typical evolution time of the MRI (see \citealt{masada07, masa08}). This is equivalent to a condition 
\begin{equation}
R_{\rm MRI} \equiv \frac{v_A^2}{\nu\Omega} \lesssim 1, \label{1}
\end{equation}
where $R_{\rm MRI}$ is the Reynolds number for the MRI, $v_A$ is the Alfv\'en velocity, $\Omega$ is the angular 
velocity and $\nu$ is the neutrino viscosity. This can be translated to the condition for the magnetic field 
\begin{eqnarray}
B \lesssim B_{\rm crit} & \equiv & (4\pi \nu\rho\Omega)^{1/2} \;, \nonumber \\
    &  = &  3.5 \times 10^{12}  \rho_{12}^{1/2} \nu_{10}^{1/2} \Omega_{2}^{1/2}\ [{\rm G}]\;, \label{eq_Bcon}
\end{eqnarray}
where $\rho_{12}$ is the density normalized by $10^{12}\ {\rm g\ cm^{-3}}$, $T_{11}$ is the temperature normalized 
by $10^{11}\ {\rm K}$, $\Omega_2$ is the angular velocity normalized by $10^2\ {\rm rad\ sec^{-1}} $, and 
$\nu$ is the neutrino viscosity normalized by $10^{10} \ {\rm cm^2\ sec^{-1}}$ (see \citealt{masada07} for the 
magnitude of the neutrino viscosity). Such enormous neutrino viscosity is plausible in the region just below the 
neutrinosphere where the strong differential rotation is developed. 

Equation~(\ref{eq_Bcon}) means that if the magnetic field is weaker than the critical value even locally, the neutrino 
viscosity suppress the linear growth of the MRI. Longaretti \& Lesur (2010) found that, in the non-resistive limit (i.e., 
highly conducting MHD fluid), the saturation level of the MRI does not be affected by the viscosity at the nonlinear 
stage despite the change of the linear growth rate. Their results might suggest that the effects of neutrino viscosity 
on the MRI-driven turbulence are only secondary in the supernova problem. 

On one hand, when the magnetic field is weaker than the critical value $B_{\rm crit}$, the linear growth of the MRI 
is suppressed and then the growth time of the fastest growing mode of the MRI $t_{\rm MRI}$ becomes
\begin{equation}
t_{\rm MRI} = \left[ \frac{2(2-q)}{q^2} \right]^{1/4} R_{\rm MRI}^{-1/2}\Omega^{-1} \;,
\end{equation} 
\citep{masa08} in contrast to that for the case with larger magnetic field than the critical value given by 
 $t_{\rm MRI} = 2/(q\Omega)$.  Figure~8 depicts the growth time of the fastest growing mode of the MRI as 
 a function of the shear rate for the cases with $B \ge B_{\rm crit}$ (solid line), $B = 10^{-2}B_{\rm crit}$ 
 (dashed line) and $B = 10^{-4}B_{\rm crit}$ (dash-dotted line). We adopt the spin rate of 
 $\Omega = 100\ {\rm rad\ sec^{-1}}$ in this figure. 
 
Considering the spin rate and shear rate plausible for the nascent PNSs [$\Omega \simeq \mathcal{O}(10^2)\ 
{\rm rad\ sec^{-1} } $ and $q \simeq 1.0$], the growth time of the MRI becomes longer than $100$ [ms], which is 
only marginally comparable to the typical timescale of neutrino-driven explosions observed in some recent 
supernova simulations (e.g., \citealt{marek,Suwa10,Takiwaki11}) when the magnetic field is much weaker than 
the critical value (dashed and dash-dotted curves). This suggests that the weaker field can not fully activate the 
MRI-driven turbulence within the dynamical timescale for the supernova explosion. From the viewpoint of the 
growth rate of the MRI as well as the saturation amplitude, we would require the large magnetic field and/or short 
spin period for the MRI-assisted supernova explosion. 

\subsection{Validity of Shearing Box Model}
As described in \citet{MRI}, the shearing box model can not treat the angular momentum transport adequately. 
In this respect, the shearing disk model is indeed superior to the shearing box model. However due to the lack of 
a simulation technique which can appropriately take into account the feed back from the global fluid motions, it may 
not be settled yet (possibly even by the semi-global simulations) whether or not the MRI will really smear out the 
differential rotation in the supernova core. It may be possible that the shear energy tapped in the differentially 
rotating layers can be supplied continuously by extracting the thermal or gravitational energies. This is considered 
to be the case in the solar interior. 

It is well known from the helioseismic observation that there exists a radial differential rotation in the upper part 
of the convectively stable radiative core (the so-called tachocline), where is suitable for the field wrapping via 
$\Omega$-effect  and the growth of the MRI (see \citealt{parfrey, masada2011}).  Although the field wrapping 
process and/or MRI might be operated in the tachocline region, the radial shear persists secularly there. The 
hydrodynamic angular momentum transport processes, such as the meridional circulation and the convective 
motion (e.g., \citealt{rudiger}), are considered to sustain the differential rotation even if the magnetic tension force 
acts to smear out the differential rotation. In the post-bounce supernova core, the convectively stable regions are 
formed between the nascent neutron star and the stalled bounce shock, in which the neutrino cooling dominates 
over the neutrino heating \citep{janka01}. We speculate that the activity of the MRI might be maintained there in 
the presence of the secular differential rotation, which is close to the situation assumed in the shearing box simulation.

To draw a robust conclusion,  one should naturally need to include the density stratification as well as the 
effects of neutrino heating/cooling which determines the convective stability. This study is only a prelude to 
improve our modeling according to the long to-do lists one by one from now on, to understand the role of MRI 
on the supernova mechanism. 

\subsection{Magnetic Field Structure during Saturation}

All our numerical models yield a turbulent, highly-tangled, magnetic field structure as the final saturated state. 
The turbulent flow and magnetic field persist during the saturation, and coherent channel structures appear 
only transiently. On the contrary, in \citet{MRI}, large-scale coherent fields with efficient angular momentum 
transport emerge after the turbulent state and are maintained for some models with a uniform initial magnetic field 
using simulation boxes of small radial and azimuthal aspect ratios $L_x/L_z$ and $L_y/L_z$. 

The magnetic field structure during saturation would be determined by whether flow-driven and current-driven 
parasitic instabilities, which are responsible for the destruction of channel solutions of the MRI, can grow or not. 
\citet{goodman} predicted analytically that these parasitic modes require radial and azimuthal wavelengths larger 
than the vertical wavelength of the channel solution for being unstable (see also \citealt{latter09}; \citealt{pessah09}). 
The coherent channel structure emerging from developed turbulence is thus expected to evolve into the large-scale 
without the destruction by the parasites in the small simulation box. 

\citet{MRI} numerically confirmed that small radial and azimuthal aspect ratios are required to maintain the large-scale 
structure of coherent magnetic fields at the saturated state (see Figs~23 and 24 in their paper). A stationary turbulent 
state with tangled magnetic fields appears even in the shearing disk model when the simulation box is large 
enough. The same trend regarding the magnetic field structure during saturation has been reported also in the shearing 
box simulation by \citet{bodo} (see also \citealt{lesaffre09}). 

As presented in \S~2, we chose the computational domain having large aspect ratios of $L_x/L_z = L_y/L_z =4$ for all 
the models. The parasitic instabilities thus can evolve without being affected by the box geometry,  
and then destroy coherent channels, yielding the less violent turbulent state continuously during the saturation. 
It should be stressed that, as studied by \citet{sano}, the recurrent formation of large-scale coherent fields can be observed 
in the saturated stage of our shearing box model when we reduce the simulation domain to that with small aspect ratios. 

When the magneto-rotational core-collapse, the MRI-active region should be confined in the upper PNS with small radial 
and latitudinal extents of $\mathcal{O}(10^5$--$10^6)$ cm (see \S~4). The azimuthal thickness of the MRI-unstable region is 
expected to be significantly larger than the radial and latitudinal ones. This geometry would inhibit the development of the coherent 
structure of magnetic fields by prompting parasitic instabilities, and lead persistently to a less violent MRI-driven turbulent state 
(cf., \citealt{MRI}). The turbulent, highly tangled magnetic field structure would be suitable for describing the MRI-active layer 
in the supernova cores. 

\section{Summary}
We performed a series of 3D compressible MHD simulation by adopting the local shearing box model.  
By changing systematically the magnitudes of the shear rate, we specifically studied how the nonlinear 
properties of the MRI-driven turbulence are controlled by the shear in the system. With applying our 
numerical results to the supernova environment, we examined the impact of the MRI on the supernova 
explosion mechanism. Our main findings are summarized as follows.

1. We could observe, in our fiducial run with $q=0.5$, three typical evolutionary stages [(i) linear 
exponential growth stage, (ii) transition stage, and (iii) nonlinear turbulent stage] which are analogous to those 
found in previous works modeling accretion disks. This validates that the variation of the shear rate does 
not change linear and nonlinear properties of the MRI qualitatively. The turbulent stress was saturated at the 
level $\langle\langle w_{\rm tot} \rangle\rangle \simeq 0.03P_0$ in our fiducial model. 

2. Our parameter survey results in the power-law relation between the shear-vorticity ratio and turbulent stress. 
The power law index $\delta$ of $\langle\langle w_{\rm tot} \rangle\rangle \propto  g_q^{\delta}$ is about 
$1/2$. The MRI-amplified magnetic energy has a similar scaling relation to the turbulent stress although the 
Maxwell stress has a power law index of $0.36$. 
 
3. It was found that the stress ratio, defined by $\langle\langle w_M \rangle\rangle/\langle\langle w_R \rangle\rangle$, 
decreases with the increase of the shear-vorticity ratio. In addition, the stress ratio calculated from our numerical results 
has the similar magnitude and $g_q$-dependence with those obtained by previous works (LKKBL09 and ZR01) despite 
the different computational settings. In the range $0.5 \lesssim q \lesssim 1.8$, the stress ratio is roughly fitted by a power 
law relation $\langle\langle w_M\rangle\rangle / \langle\langle w_R\rangle\rangle \propto g_q^{-3/4}$. 
 
4. The stronger magnetic field or the larger shear rate provides the higher energy release due to the MRI-driven 
turbulence. For a rapidly rotating PNS with the spin period in milliseconds and with strong magnetic fields of
$10^{15}$ G, the energy dissipation rate is estimated to exceed $10^{51}\ {\rm erg\ sec^{-1}}$ . Our results suggest 
that the conventional MHD mechanisms of core-collapse supernovae are likely to be affected by the MRI-driven turbulence, 
which we speculate, on one hand, could harm the MHD-driven explosions due to the dissipation of the shear rotational 
energy at the PNS surface, on the other hand the energy deposition there might be potentially favorable for the working 
of the neutrino-heating mechanism.

\acknowledgments
We thank the anonymous referee for constructive comments. 
Numerical computations were carried on SX8 of Institute of Laser Engineering, Osaka University, and partly on XT4 and 
general common use computer system at the center for Computational Astrophysics, CfCA, the National Astronomical Observatory of 
Japan.  This study was supported in part by the Grants-in-Aid for the Scientific Research from the Ministry of Education, Science and 
Culture of Japan (Nos. 19104006, 19540309, 20740150, 23540323, and 24740125) and by HPCI Strategic Program of Japanese 
MEXT. 

\clearpage
\begin{figure}[tp]
\begin{center}
\scalebox{0.4}{{\includegraphics{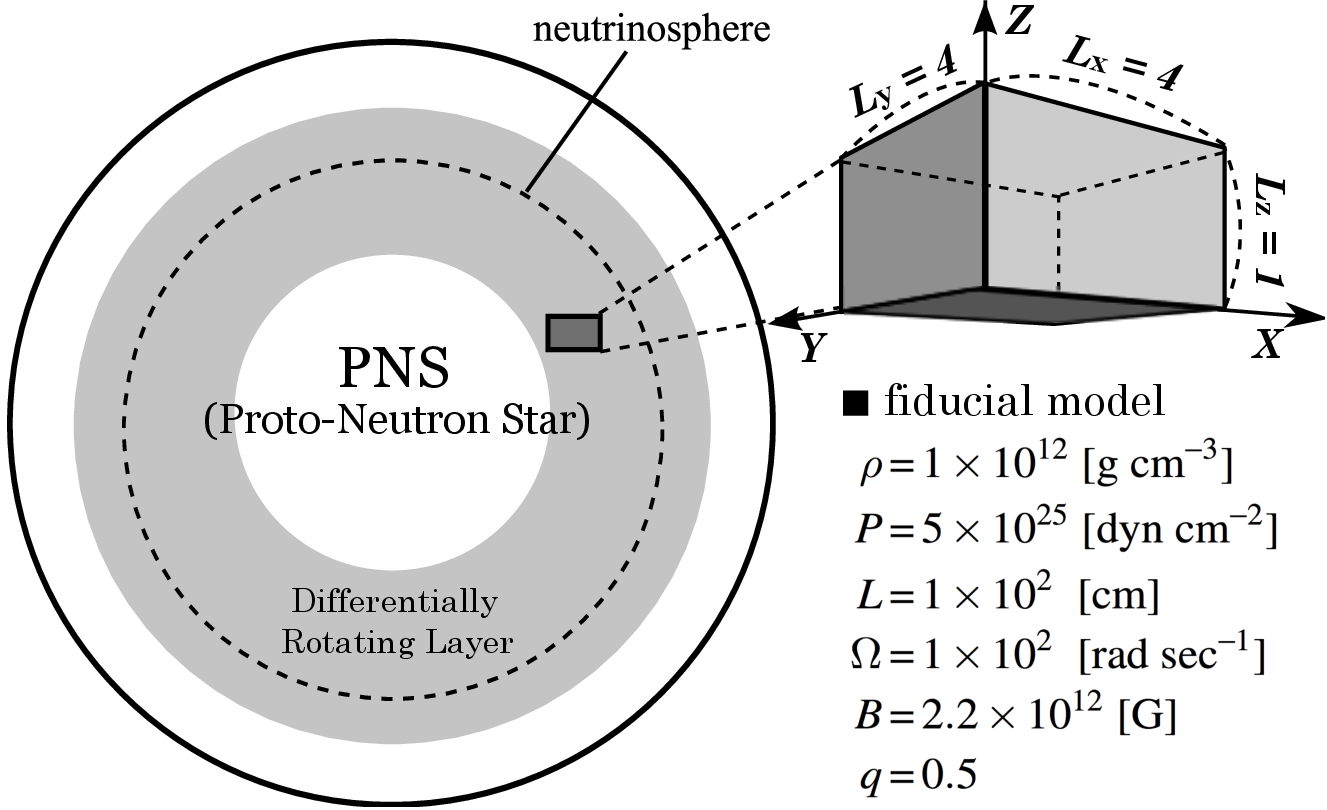}}} 
\caption{Schematic picture of the numerical setting in our calculations. The differentially rotating, neutrino-thick 
region below the neutrinosphere is focused ($\simeq $ gray shaded region). The temporal evolution of the 
the local portion of the upper proto-neutron stars is studied by the local shearing box model. The physical 
parameters adopted for our fiducial run are summarized in this figure. }
\label{fig1}
\end{center}
\end{figure}
\begin{figure}[tp]
\begin{center}
\scalebox{1.0}{{\includegraphics{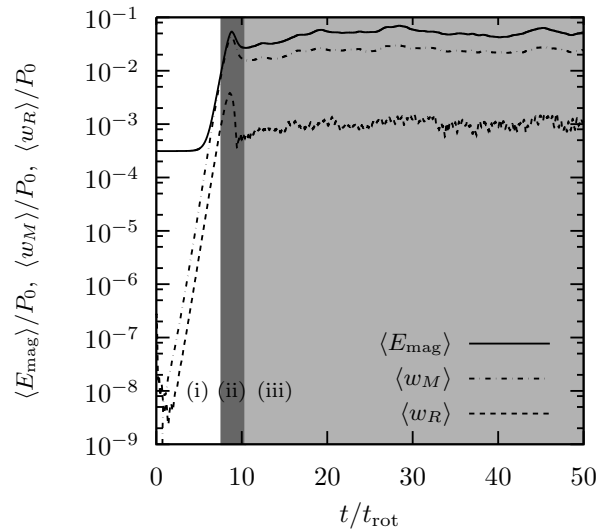}}} 
\caption{Temporal evolution of volume-averaged magnetic energy (solid), Maxwell stress (dash-dotted), and 
Reynolds stress (dashed) for the fiducial model with $q=0.5$. The vertical and horizontal axes are normalized 
by initial gas pressure $P_0$ and rotation period $t_{\rm rot}$ respectively. Three typical evolutionary stages,  
that is (i) linear exponential growth stage, (ii) transition stage, and (iii) nonlinear turbulent stage are filled 
by white, dark gray and light gray shaded colors.}
\label{fig2}
\end{center}
\end{figure}
\begin{figure*}[tbp]
\begin{center}
{\scalebox{0.55}{{\includegraphics{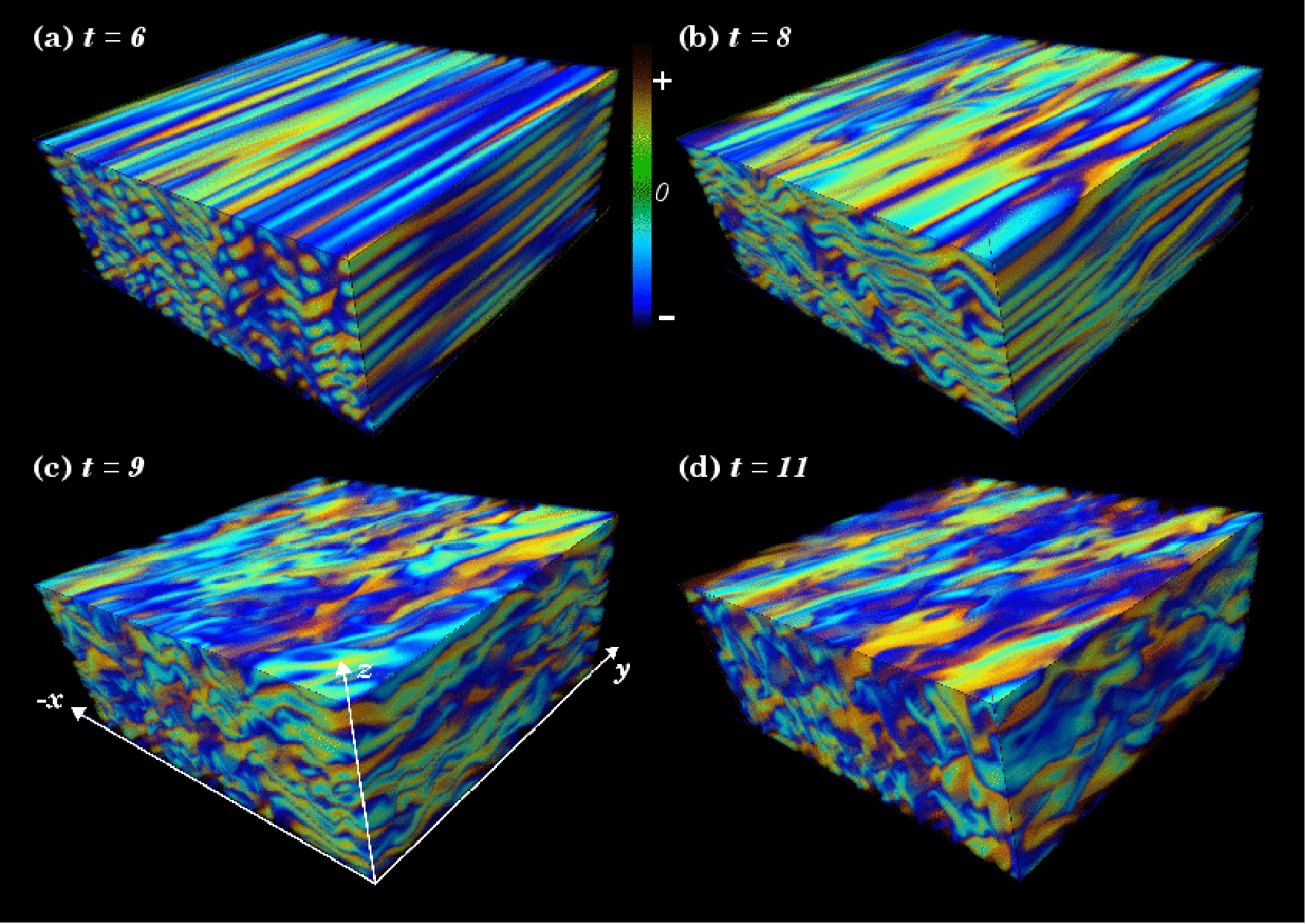}}} }
\caption{3D Volume rendering of the radial magnetic field for the fiducial model at the times 
(a) $t=6t_{\rm rot}$, (b) $t=8t_{\rm rot}$, (c) $t=10t_{\rm rot}$, and (d) $t=11t_{\rm rot}$. }
\label{fig3}
\end{center}
\end{figure*}
\begin{figure}[bp]
\begin{center}
\scalebox{1.0}{{\includegraphics{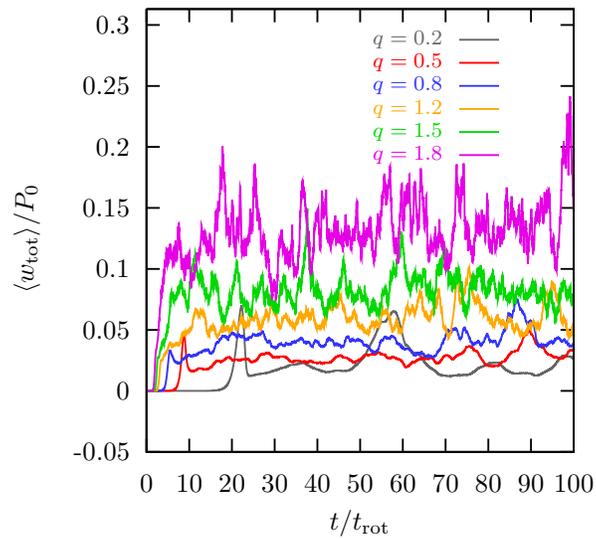}}} 
\caption{Temporal evolution of volume-averaged total stress $\langle w_{\rm tot} \rangle$ 
normalized by the initial gas pressure $P_0$ for the models with different shear rates of $q=0.2$ (grey), 
$0.5$ (red), $0.8$ (blue), $1.2$ (orange), $1.5$ (green) and $1.8$ (purple) respectively. }
\label{fig4}
\end{center}
\end{figure}
\begin{figure*}[tpb]
\begin{center}
\scalebox{0.6}{{\includegraphics{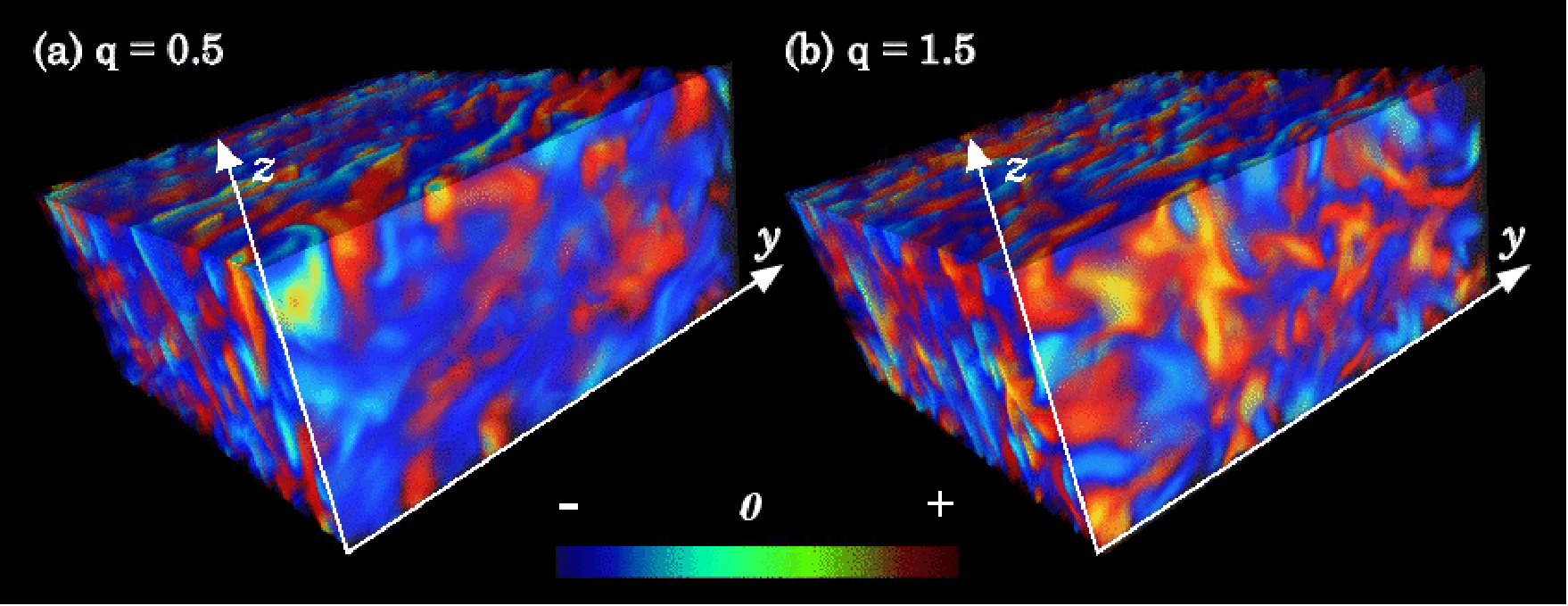}}} 
\caption{3D volume rendering of the radial magnetic field for the models with (a) $q=0.5$, and 
(b) $q=1.5$. The color bar indicates the amplitude of the radial magnetic field, that is the red 
denotes the positive value and the blue is the negative.}
\label{fig5}
\end{center}
\end{figure*}
\begin{figure*}[tbp]
\begin{center}
\begin{tabular}{cc}
\scalebox{1.0}{{\includegraphics{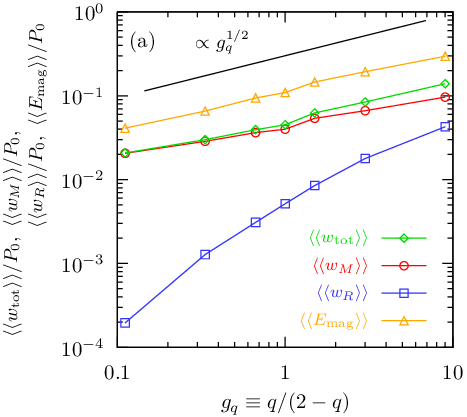}}} &
\scalebox{1.0}{{\includegraphics{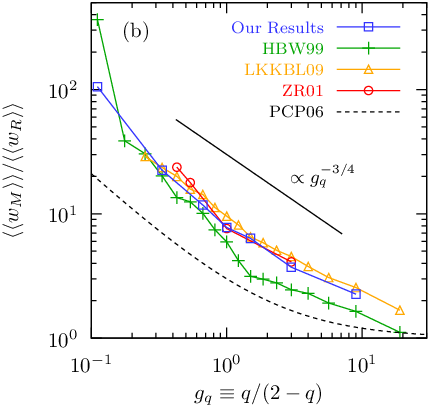}}} 
\end{tabular}
\caption{(a). The volume- and time-averaged Maxwell (red circles), Reynolds (blue squares) and total stresses 
(green diamonds) as a function of the shear-vorticity ratio $g_q$. The yellow triangles represent time- and 
volume- averaged magnetic energy at the saturated state. (b). The ratio of Maxwell and Reynolds stresses for the 
models with different initial settings as a function of the shear-vorticity ratio $g_q$. The blue squares, green crosses, 
yellow triangles, and red circles demonstrate the results of our work, Hawley et al. (1999), Liljestr\"om et al. (2009), 
and Ziegler\&Rudiger (2001) respectively. The dashed line traces the prediction from the local model developed 
by Pesah et al. (2006). The solid line gives a power law relation of $\langle\langle w_M\rangle\rangle / \langle\langle 
w_R\rangle\rangle \propto g_q^{-3/4}$. }
\label{fig6}
\end{center}
\end{figure*}
\begin{figure*}[tbp]
\begin{center}
\begin{tabular}{cc}
\scalebox{1.0}{{\includegraphics{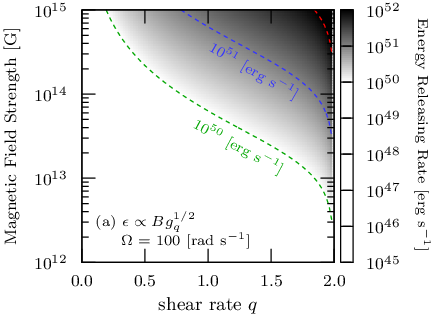}}} &
\scalebox{1.0}{{\includegraphics{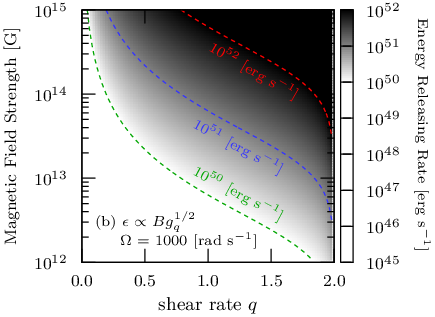}}} \\
\scalebox{1.0}{{\includegraphics{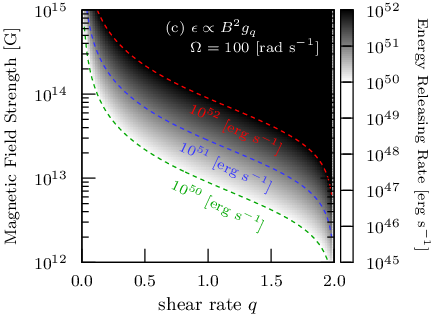}}} &
\scalebox{1.0}{{\includegraphics{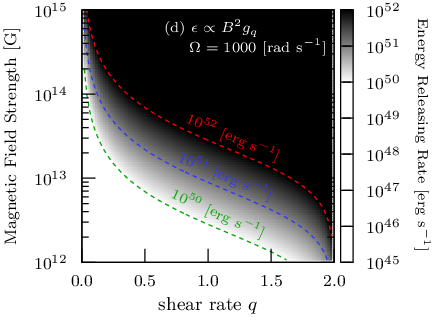}}} \\
\end{tabular}
\caption{The energy releasing rate due to the MRI turbulence $L_{\rm MRI}$ as functions of the magnetic field strength 
$B$ and the shear rate $q$. Panels~(a) and (b) show the models with the weakest parameter dependence of $\xi =1$ 
and $\delta = 1/2$, while Panels~(c) and (d) are for the models with the strongest parameter dependence of $\xi =2$ 
and $\delta = 1$. The slow rotation ($\Omega = 100\ {\rm rad\ sec^{-1}}$) is assumed in left panels and the rapid 
rotation ($\Omega = 1000\ {\rm rad\ sec^{-1}}$) is in right panels. The green, blue and red dashed curves correspond 
to the energy releasing rates $10^{50}$, $10^{51}$, and $10^{52}\ {\rm erg\ sec^{-1}}$ 
respectively. Note that the darker color gives the higher energy releasing rate.}
\label{fig7}
\end{center}
\end{figure*}
\begin{figure}[tbp]
\begin{center}
\scalebox{1.0}{{\includegraphics{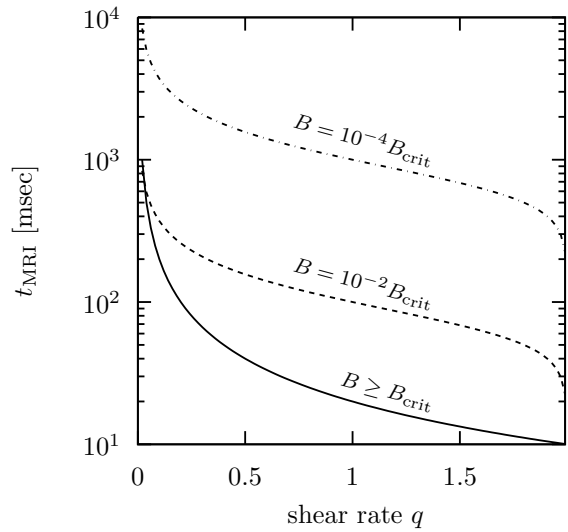}}} 
\caption{The typical linear growth time of the MRI as a function of the shear rate for the cases with 
$B \ge B_{\rm crit}$ (solid line), $B = 10^{-2}B_{\rm crit}$ (dashed line)  and $B = 10^{-4}B_{\rm crit}$ 
(dash-dotted line), where $B_{\rm crit} = 3.5 \times 10^{12}  \rho_{12}^{1/2} \nu_{10}^{1/2} 
\Omega_{2}^{1/2}\ [{\rm G}]$. We adopt  the moderate spin rate of $\Omega = 100\ {\rm rad\ sec^{-1}}$ 
in this figure.}
\label{fig8}
\end{center}
\end{figure}
\begin{table*}[tpb]
\begin{center}
\begin{tabular}{@{}cccccc}\hline\hline
& shear rate $q$ & $\langle\langle w_{M}\rangle\rangle /P_0$ &  $\langle\langle w_{R}\rangle\rangle/P_0$ &  $\langle\langle E_{\rm mag}\rangle\rangle/P_0$  & $\langle\langle w_{M}\rangle\rangle /\langle\langle w_{R}\rangle\rangle $ \\
\hline\hline
Model1 & 0.2 & $2.06 \times 10^{-2}$ & $1.96 \times 10^{-4}$ & $4.11 \times 10^{-2}$ & $1.05 \times10^{2}$ \\
Model2 & 0.5 & $2.86 \times 10^{-2}$ & $1.28 \times 10^{-3}$ & $6.57 \times 10^{-2}$ & $2.23 \times10^{1}$\\
Model3 & 0.8 & $3.63 \times 10^{-2}$ & $3.09 \times 10^{-3}$ & $9.41 \times 10^{-2}$ & $1.18 \times10^{1}$\\
Model4 & 1.0 & $3.99 \times 10^{-2}$ & $5.15 \times 10^{-3}$ & $1.09 \times 10^{-1}$ & $7.78 \times10^{0}$\\
Model5 & 1.2 & $5.38 \times 10^{-2}$ & $8.49 \times 10^{-3}$ & $1.46 \times 10^{-1}$ & $6.33 \times10^{0}$\\
Model6 & 1.5 & $6.64 \times 10^{-2}$ & $1.78 \times 10^{-2}$ & $1.94 \times 10^{-1}$ & $3.72 \times10^{0}$\\
Model7 & 1.8 & $9.63 \times 10^{-2}$ & $4.27 \times 10^{-2}$ & $2.96 \times 10^{-1}$ & $2.26 \times10^{0}$\\

\hline\hline
\end{tabular}
\caption{The shear rates used in our parametric survey are listed in the first column. The second, third, and forth 
column show the time- and volume-averaged Maxwell stress  $\langle\langle w_{M}\rangle\rangle $, Reynolds 
stress $\langle\langle w_{R}\rangle\rangle $, and Magnetic energy  $\langle\langle E_{mag}\rangle\rangle $ 
normalized by the initial gas pressure $P_0 = 5\times 10^{-7}$. The ratio of Maxwell and Reynolds stresses 
$\langle\langle w_{M}\rangle\rangle /\langle\langle w_{R}\rangle\rangle $ is in the fifth column. 
The time-average is taken over the saturated stage from $50t_{\rm rot}$ to $150t_{\rm rot}$ for models 1,2,3, and 
from $100t_{\rm rot}$ to $150t_{\rm rot}$ for models 4,5,6,7. Model 2 corresponds to the fiducial one.} 
\label{table1}
\end{center}
\end{table*}

\end{document}